# Link Mining for Kernel-based Compound-Protein Interaction Predictions Using a Chemogenomics Approach

Masahito Ohue[1,2,3,4*], Takuro Yamazaki[3], Tomohiro Ban[4], and Yutaka Akiyama[1,2,3,4*]

[1]Department of Computer Science, School of Computing,
Tokyo Institute of Technology, Japan
[2]Advanced Computational Drug Discovery Unit, Institute of Innovative Research,
Tokyo Institute of Technology, Japan
[3]Department of Computer Science, Faculty of Engineering,
Tokyo Institute of Technology, Japan
[4]Department of Computer Science, Graduate School of Information Science and Engineering,
Tokyo Institute of Technology, Japan

*ohue@c.titech.ac.jp,  akiyama@c.titech.ac.jp

**Abstract.** Virtual screening (VS) is widely used during computational drug discovery to reduce costs. Chemogenomics-based virtual screening (CGBVS) can be used to predict new compound-protein interactions (CPIs) from known CPI network data using several methods, including machine learning and data mining. Although CGBVS facilitates highly efficient and accurate CPI prediction, it has poor performance for prediction of new compounds for which CPIs are unknown. The pairwise kernel method (PKM) is a state-of-the-art CGBVS method and shows high accuracy for prediction of new compounds. In this study, on the basis of link mining, we improved the PKM by combining link indicator kernel (LIK) and chemical similarity and evaluated the accuracy of these methods. The proposed method obtained an average area under the precision-recall curve (AUPR) value of 0.562, which was higher than that achieved by the conventional Gaussian interaction profile (GIP) method (0.425), and the calculation time was only increased by a few percent.

## 1    Introduction

Virtual screening (VS), in which drug candidate compounds are selected by a computational method, is one of the main processes in the early stages of drug discovery. There are three main approaches to VS: ligand-based VS (LBVS) [1] using known activity information for the target protein of the drug; structure-based VS (SBVS) [2] using structural information for the target protein of the drug; and chemogenomics-based VS (CGBVS) [3] based on known interaction information for multiple proteins and multiple compounds (also called drug-target interaction prediction). Both LBVS and CGBVS do not require a protein tertiary structure, and both depend on statistical machine learning using known experimental activity data. However, CGBVS yields more robust prediction results by handling multiple types of proteins. CGBVS has been well studied in recent years [4][5][6][7][8][9], and a review of CGBVS was recently published by Ding *et al.* [3].

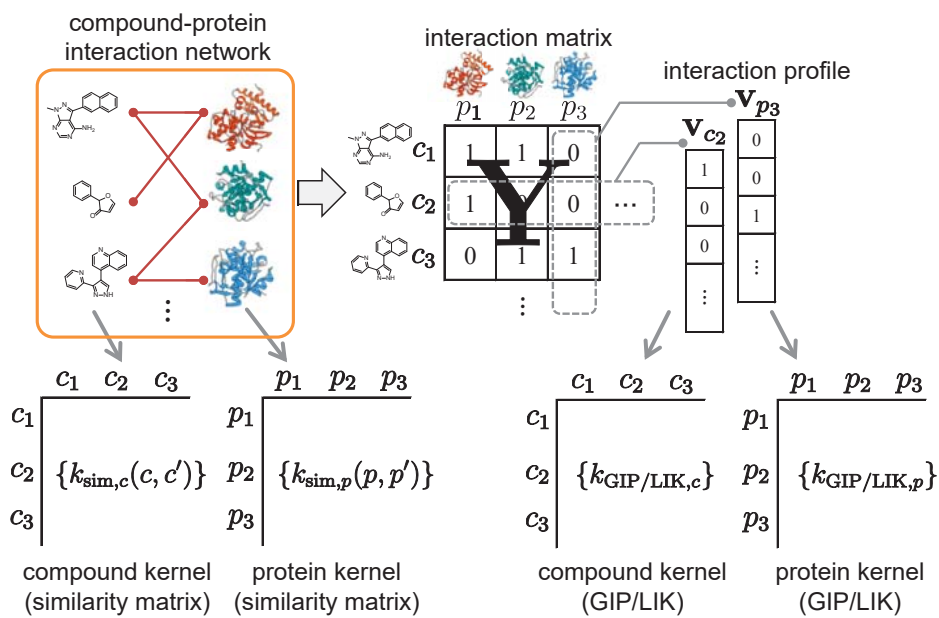

**Fig. 1.** Schematic diagram of information used in CGBVS. Information on interaction matrices and interaction profiles can be obtained from CPI data. Kernel matrices were obtained from the relationship between the compounds and the proteins.

In CGBVS, computations are mainly performed using a similarity matrix of proteins, a similarity matrix of compounds, and an interaction profile matrix composed of binary values with and without interactions (Fig. 1).

The kernel method is often applied for prediction [5][6][7]. Conventionally, in CGBVS, an interaction profile matrix is used only as labeled training data; however, an increasing number of frameworks have recently been described that utilize interaction profile matrices as their main features. The Gaussian interaction profile (GIP) is one of these frameworks [5]. In addition to information regarding similarity matrices, GIP uses similarities between vectors when an interaction profile matrix is viewed as vertical and horizontal vectors. The GIP kernel functions of proteins and compounds are represented as follows (details are described in Section 2):

$$k_{\text{GIP},c}(c,c') = \exp\left(-\gamma_c \left\|\mathbf{v}_c - \mathbf{v}_{c'}\right\|^2\right), \quad k_{\text{GIP},p}(p,p') = \exp\left(-\gamma_p \left\|\mathbf{v}_p - \mathbf{v}_{p'}\right\|^2\right). \quad (1)$$

The problem with GIP kernels is that '0' bit (interaction is unknown) is taken into account, similarly to '1' bit (interaction). Thus, $k_{\text{GIP}}$ shows a maximum value (which is equivalent to two compounds with common interaction partners) for two novel compounds when the interactions with all proteins are unknown. Since '0' potentially includes both no interactions and unknown interactions in general CGBVS problems and benchmark datasets, information for '1' should be considered more reliable.

In this way, as a framework that mainly considers '1' rather than '0', link mining has emerged for calculation of links within a network. Link mining is a framework applied to analyze networks such as social networks and the World Wide Web. Nodes and edges (links) of a network are used in the calculation. If nodes of the network are proteins/compounds, and edges are drawn in the interacting compound-protein pair, analysis using the framework of link mining becomes possible. Some reports have also applied the method of link mining directly to the problem of CGBVS [8][9]. However, these methods do not use the framework of the kernel method.

Therefore, in this study, we propose to use the link indicator kernel (LIK), based on link indicators used in link mining, as the kernel with an interaction profile matrix such as that formed from GIP kernels. According to a review by Ding *et al.*, the pairwise kernel method (PKM) [7] using a support vector machine (SVM) as a kernel learning scheme is superior in learning performance to CGBVS [3]. Thus, we integrated GIP and LIK kernels to the PKM and showed that LIK kernels could capture the effects of interaction profiles.

## 2      Materials and Methods

### 2.1      Preliminary

An overview of the compound-protein interaction prediction problem is shown in Figure 1. Similarities were defined between compounds and between proteins, with the Tanimoto coefficient of fingerprints (e.g. ECFP [10], SIMCOMP [11]) or Euclid distance of physicochemical properties for compounds, and the Euclid distance of $k$-mer amino acid sequence profiles or Smith-Waterman alignment scores [12] for proteins.

The interaction $y(c, p)$ between a compound $c$ and a protein $p$ is defined as binary {0, 1}, where '1' represents an interaction (e.g., $c$ is the active compound for the protein $p$) and '0' represents no interaction (often including unknowns). For learning, the $n_c \times n_p$ matrix $\mathbf{Y} = \{y(c, p)\}_{c, p}$ (called the interaction matrix) was used as training data, where $n_c$ is the number of target compounds and $n_p$ is the number of target proteins. Interactions were predicted for pairs of compounds and proteins using the learned model. When looking at each row and each column of the interaction matrix as a vector, the vector was called the interaction profile. The interaction profile $\mathbf{v}_c$ of compound $c$ was $\mathbf{v}_c = (y(c, p_1), y(c, p_2), ..., y(c, p_{np}))^\text{T}$, and the interaction profile $\mathbf{v}_p$ of protein $p$ was $\mathbf{v}_p = (y(c_1, p), y(c_2, p), ..., y(c_{nc}, p))^\text{T}$.

### 2.2    PKM

The pairwise kernel method (PKM) [7] developed by Jacob *et al.* is based on pairwise kernels and tensor product representation for compound and protein vectors. Normally, a map $\Phi(c, p)$ for a pair of compounds and proteins $(c, p)$ is required for a learning scheme. In the PKM, the learning scheme involves the tensor product of the map of compound $\Phi_c(c)$ and the map of protein $\Phi_p(p)$. Therefore, $\Phi(c, p)$ is represented as follows:

$$\Phi(c, p) = \Phi_c(c) \otimes \Phi_p(p), \tag{2}$$

where $\otimes$ is the tensor product operator. Pairwise kernel $k$ is defined between two pairs of proteins and compounds $(c, p)$ and $(c', p')$ as follows:

$$\begin{aligned} k((c, p), (c', p')) &\equiv \Phi(c, p)^T \Phi(c', p') \\ &= (\Phi_c(c) \otimes \Phi_p(p))^T (\Phi_c(c') \otimes \Phi_p(p')) \\ &= \Phi_c(c)^T \Phi_c(c') \times \Phi_p(p)^T \Phi_p(p') \\ &\equiv k_c(c, c') \times k_p(p, p') , \end{aligned} \tag{3}$$

where $k_c$ is a compound kernel between two compounds, and $k_p$ is a protein kernel between two proteins. Thus, it is possible to find the kernel $k$ between two compound-protein pairs with the scalar product of $k_c$ and $k_p$. If both $k_c$ and $k_p$ are positive definite kernels, $k$ is also a positive definite kernel. The similarity value mentioned in Section 2.1 is often used for $k_c$ and $k_p$ [4].

### 2.3    GIP

The Gaussian interaction profile (GIP) [5] was developed to incorporate interaction profiles into kernel learning by van Laarhoven *et al*. The GIP kernel $k_{\text{GIP}}$ based on the radial basis function shown in Eq. (1) is used for the compound kernel $k_c$ and protein kernel $k_p$ in Eq. (3). Here,

$$\gamma_c = \left( \frac{1}{n_c} \sum_{i=1}^{n_c} \| \mathbf{v}_{c_i} \|^2 \right)^{-1}, \quad \gamma_p = \left( \frac{1}{n_p} \sum_{i=1}^{n_p} \| \mathbf{v}_{p_i} \|^2 \right)^{-1}. \tag{4}$$

Importantly, $k_{\text{GIP}}$ is never used alone for $k_c$ and $k_p$, but is used as a multiple kernel (simple weighted average) in combination with similarity-based kernels:

$$\begin{aligned} k_c(c, c') &= k_{\text{sim},c}(c, c') + w_{k,c} k_{\text{GIP},c}(c, c') \\ k_p(p, p') &= k_{\text{sim},p}(p, p') + w_{k,p} k_{\text{GIP},p}(p, p') , \end{aligned} \tag{5}$$

where $w_{k,c}$ and $w_{k,p}$ are weighted parameters for multiple kernels.

5     M. Ohue, T. Yamazaki, T. Ban, and Y. Akiyama## 2.4 LIK

The link indicator is an index used for network structural analysis, such as analysis of the hyperlink structure of the World Wide Web and friend relationships in social network services. In this study, we proposed link indicator kernels (LIKs) based on the link indicators for compound-protein interaction networks to incorporate interaction profiles into kernel learning. We selected three link indicators:

$$\textit{Jaccard index} \quad k_{\text{LIK-Jac}}(\mathbf{v}, \mathbf{v}') = \frac{\mathbf{v}^\text{T} \mathbf{v}'}{\|\mathbf{v}\|^2 + \|\mathbf{v}'\|^2 - \mathbf{v}^\text{T} \mathbf{v}'} \tag{6}$$

$$\textit{Cosine similarity} \quad k_{\text{LIK-cos}}(\mathbf{v}, \mathbf{v}') = \frac{\mathbf{v}^\text{T} \mathbf{v}'}{\|\mathbf{v}\| \|\mathbf{v}'\|} \tag{7}$$

$$\textit{LHN} \quad k_{\text{LIK-LHN}}(\mathbf{v}, \mathbf{v}') = \frac{\mathbf{v}^\text{T} \mathbf{v}'}{\|\mathbf{v}\|^2 \|\mathbf{v}'\|^2} . \tag{8}$$

These link indicators become positive definite kernels when used as kernels. Cosine similarity and LHN are positive definite kernels because of the properties of the kernel function[1] and the positive definite of the inner product between the two vectors, and the Jaccard index was previously proven to be positive definite by Bouchard *et al.* [13]. There are other link indicators, such as the Adamic-Adar index and graph distance; however, because these are not positive definite kernels, as required for kernel methods, they were not used in this study.

For integration of LIK and PKM (similarity kernels), the same method applied for GIP was adopted. That is, considering multiple kernels, the kernels were defined as:

$$\begin{aligned} k_c(c,c') &= k_{\text{sim},c}(c,c') + w_{k,c} k_{\text{LIK},c}(c,c') \\ k_p(p,p') &= k_{\text{sim},p}(p,p') + w_{k,p} k_{\text{LIK},p}(p,p') \end{aligned}, \tag{9}$$

where $k_{\text{LIK},c}$ and $k_{\text{LIK},p}$ are LIKs for two compounds and two proteins, respectively.

## 2.5 Implementation

In this study, we used scikit-learn [14], a Python library for machine learning, to implement PKM, GIP, and LIK. As a kernel learning method, SVM can be used for scikit-learn based on LIBSVM [15]. For the link indicator calculation of LIK, we used the python library networkx [16].

---

[1] Let $k: X \times X \to \mathbb{R}$ be a positive definite kernel and $f: X \to \mathbb{R}$ be an arbitrary function. Then, the kernel $k'(\mathbf{x}, \mathbf{y}) = f(\mathbf{x}) k(\mathbf{x}, \mathbf{y}) f(\mathbf{y})$ $(\mathbf{x}, \mathbf{y} \in X)$ is also positive definite.

### 2.6   Dataset and Performance Evaluation

We used the benchmark dataset of CPI predictions published by Yamanishi *et al*. [4] according to the review of Ding *et al*. [3]. It is a well-known and well-used benchmark dataset in the field. The dataset consisted of four target protein groups ("Nuclear Receptor", "GPCR", "Ion Channel", and "Enzyme"). The SIMCOMP score [11] was used for compound similarity, and the normalized Smith-Waterman score [12] was used for protein similarity, as calculated by Yamanishi *et al*. [4]. Information on the interaction matrix was also provided by Yamanishi *et al*. [4]. Evaluation was performed by cross validation (CV). Three types of CVs were tested: randomly selected from all compound-protein pairs (pairwise CV), randomly selected compounds (compound CV), and randomly selected proteins (protein CV). The outlines of these CVs are shown in Figure 2. According to Ding *et al*. [3], the area under the receiver operating characteristic curve (AUROC) and the area under the precision-recall curve (AUPR) were calculated for the evaluation value of 10-fold CVs. Each accuracy value was averaged five times for 10-fold CVs with different random seeds. Note that the cost parameter $C$ of SVM was optimized from {0.1, 1, 10, 100} in 3-fold CVs according to Ding *et al*. [3]. The multiple kernel weights $w_k$ of Eqs. (5) and (9) have the same values for proteins and compounds, and we evaluated {0.1, 0.3, 0.5, 1}.

## 3   Results and Discussion

### 3.1   Performance of the Proposed Method for Cross-Validation Benchmarking

Figure 3 shows the results for the prediction accuracy of the average values of three types of CVs in the four Yamanishi datasets (i.e., average values of 12 prediction accuracy values). We tested multiple kernel weights $w_k$ in four patterns, and LIK with cosine similarity was the most accurate for both AUPR and AUROC (AUPR: 0.562 and AUROC: 0.906). In the case of cosine similarity, the weight $w_k = 0.5$ showed the best performance. Compared with GIP, the prediction accuracy of LIK showed higher accuracy overall.

**compound-wise CV**

|       | $p_1$ | $p_2$ | $p_3$ | $p_4$ | $p_5$ | $p_6$ |
|-------|---|---|---|---|---|---|
| $c_1$ | 1 | 0 | 0 | 1 | 0 | 0 |
| $c_2$ | 0 | 1 | 0 | 0 | 1 | 1 |
| $c_3$ | 1 | 0 | 0 | 0 | 0 | 0 |
| $c_4$ | 1 | 0 | 1 | 1 | 0 | 0 |
| $c_5$ | ? | ? | ? | ? | ? | ? |
| $c_6$ | ? | ? | ? | ? | ? | ? |

**protein-wise CV**

|       | $p_1$ | $p_2$ | $p_3$ | $p_4$ | $p_5$ | $p_6$ |
|-------|---|---|---|---|---|---|
| $c_1$ | 1 | 0 | 0 | 1 | ? | ? |
| $c_2$ | 0 | 1 | 0 | 0 | ? | ? |
| $c_3$ | 1 | 0 | 0 | 0 | ? | ? |
| $c_4$ | 1 | 0 | 1 | 1 | ? | ? |
| $c_5$ | 0 | 1 | 0 | 0 | ? | ? |
| $c_6$ | 0 | 0 | 1 | 1 | ? | ? |

**pairwise CV**

|       | $p_1$ | $p_2$ | $p_3$ | $p_4$ | $p_5$ | $p_6$ |
|-------|---|---|---|---|---|---|
| $c_1$ | ? | 0 | ? | 1 | 0 | ? |
| $c_2$ | 0 | ? | 0 | ? | 1 | 1 |
| $c_3$ | 1 | 0 | 0 | 0 | 0 | ? |
| $c_4$ | 1 | 0 | ? | ? | 0 | 0 |
| $c_5$ | ? | 1 | 0 | 0 | 0 | 1 |
| $c_6$ | ? | 0 | ? | 1 | ? | 0 |

**Fig. 2.** Conceptual diagram of three types of cross-validations (CVs): compound-wise CV, protein-wise CV, and pairwise CV. A case with a 3-fold CV is shown as an example. The "?" indicates samples to be used for the test set.

7     M. Ohue, T. Yamazaki, T. Ban, and Y. Akiyama

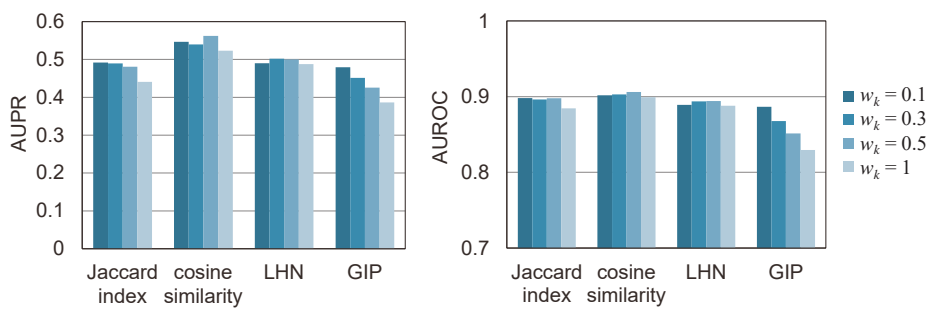

**Fig. 3.** Overall prediction accuracy for each CPI prediction method in 10-fold CV tests. The AUPR and AUROC values are averaged values of three types of CVs and four types of datasets (total average for 12 AUPR/AUROC values). For 10-fold CVs, calculations were performed five times with different random seeds, and the accuracy values were then averaged.

Figure 4 shows the mean value of the prediction accuracy for each CV, including compound-wise, protein-wise, and pairwise CVs. Division of the dataset in each CV was randomly tried five times, and the values were averaged. The multiple kernel weight $w_k$ was set to the best value in the cross-validation results (shown in Figure 3). In the evaluation of AUROC, GIP showed accuracy comparable to that of the three LIKs; however, similar results were not observed for AUPR. In particular, we found that LIK showed a high value in the AUPR evaluation for compound-wise and protein-wise CVs, which could be evaluated for predictive performance for novel compounds and proteins, respectively.

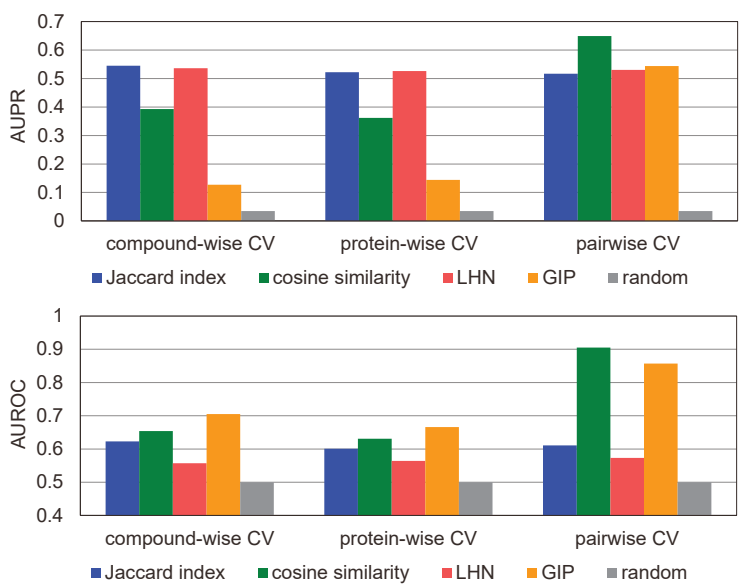

**Fig. 4.** Prediction accuracy of protein-wise, compound-wise, and pairwise CVs. In random prediction cases, an AUROC value of 0.5 and an AUPR value of 0.035 were obtained (averaged values depending on the ratio of positive samples on the dataset).

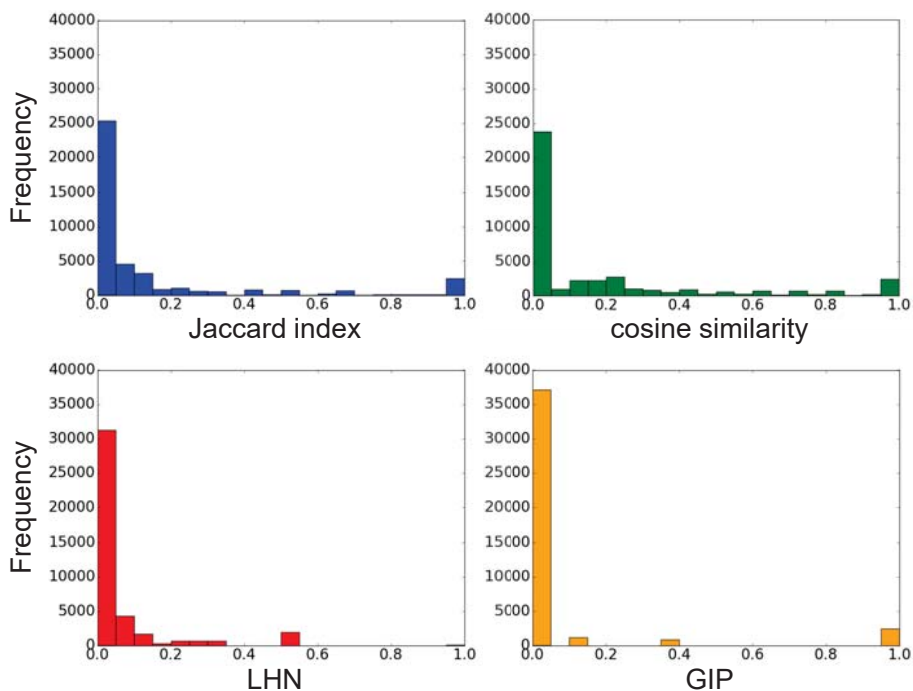

**Fig. 5.** Distribution of values of LIKs (Jaccard index, cosine similarity, and LHN) and GIP given all protein interaction profiles in the Yamanishi dataset.

### 3.2 Observed Distribution of Link Indicator Frequency

Distribution of values of four protein interaction profile similarities ($k_{LIK}(p, p')$) were calculated using each link indicator to determine why LIKs with cosine similarity showed better results. The histograms of similarity values are shown in Figure 5. From this result, the Jaccard index and LHN were found to have relatively similar distributions of similarity values between 0 and 1 (i.e., the intermediate value was low). Additionally, the number of pairs whose similarities ranged from 0.95 to 1 had the highest cosine similarity. This may be related to the AUROC value of cosine similarity, which tended to be higher. Conversely, for LHN, which showed the lowest similarity from 0.95 to 1, precision may be higher, and AUPR may tend to be higher. GIP also consisted of a few intermediate values similarly to LHN. Overall, cosine similarity showed the best performance in this study. A gentle distribution using an intermediate value may be more effective as information for the compound-protein network structure.

### 3.3 Computational Complexity

The computational complexity for constructing the prediction model for the PKM is $O(n_c^3 n_p^3)$. However, the computational complexity for calculating the link indica-tors used in this study was $O(n_c n_p(n_c + n_p))$. Thus, the computational complexity of our proposed method was $O(n_c^3 n_p^3 + n_c n_p(n_c + n_p))$. Here, $n_c$ and $n_p$ were greater than 1 in general, and thus, $n_c^3 n_p^3$ was greater than $n_c n_p(n_c + n_p)$. Therefore, the computational complexity was $O(n_c^3 n_p^3)$, which was the same as those of PKM and GIP. Our method could predict CPIs without a major increase in calculation time. The execution time of one run of 10 runs of 10-fold CVs is shown in Table 1. The results in the table are shown for computations running on an ordinary personal computer with an Intel Core i5 CPU. Thus, our proposed method showed a slight increase in the execution time by several percentage points as compared with that of PKM ("Nuclear Receptor" had the highest rate of increase due to the small dataset).

### 3.4  Limitations and Challenges

The proposed method can be directly applied to prediction based on the GIP (e.g., WNNGIP [6], KBMF2K [17], and KronRLS-MKL [18]), and improvement of prediction accuracy is expected. For example, WNNGIP can provide robust predictions for compounds and proteins with less interaction information by complementing the interaction matrix with the weighted nearest neighbor method in advance. However, kernel-based methods, including the proposed method, are restricted to the framework using kernel functions. For example, it is not possible to simply combine LIK or GIP with the method based on matrix factorization (e.g., NRLMF [19]). Further mathematical ideas and computational experiments are needed to develop integrated methods.

## 4    Conclusions

In this study, we proposed a kernel method using link indicators from the viewpoint of link mining to utilize the information of the CPI network for machine learning. We attempted to utilize three link indicators (Jaccard index, cosine similarity, and LHN) for construction of positive definite kernels and compared them with the GIP method when combined with the SVM-based PKM method. As a result, learning by multiple kernels using LIK with cosine similarity and setting of the kernel weight $w_k$ to 0.5 showed the best prediction accuracy (averaged AUPR = 0.562). In both AUROC and AUPR, the improvement of LIK accuracy was confirmed compared with that of GIP.

**Table 1.** Comparison of calculation times for the PKM and proposed method in each dataset. The time taken to calculate one time out of 10 calculations of 10-fold CVs is shown.

|  | Nuclear Receptor | GPCR | Ion Channel | Enzyme |
|---|---|---|---|---|
| Conventional (PKM) [sec] | 0.0680 | 4.86 | 24.1 | 232 |
| Proposed (PKM plus LIK) [sec] | 0.0850 | 5.17 | 24.8 | 239 |
| Increase rate (%) | 25 | 6.4 | 2.9 | 3.3 |

**Acknowledgments**. This work was partially supported by the Japan Society for the Promotion of Science (JSPS) KAKENHI (grant numbers 24240044 and 15K16081), and Core Research for Evolutional Science and Technology (CREST) "Extreme Big Data" (grant number JPMJCR1303) from the Japan Science and Technology Agency (JST).